\documentstyle[preprint,aps]{revtex}
\newcommand{\bmp}{{\mbox{\boldmath $p$}}}
\newcommand{\bms}{{\mbox{\boldmath $s$}}}
\newcommand{\bmP}{{\mbox{\boldmath $P$}}}
\newcommand{\bmr}{{\mbox{\boldmath $r$}}}
\newcommand{\bmz}{{\mbox{\boldmath $r_1$}}}
\newcommand{\bmzp}{{\mbox{\boldmath $r_1'$}}}

\newcommand{\bmq}{{\mbox{\boldmath $q$}}}
\newcommand{\bmb}{{\mbox{\boldmath $b$}}}
\begin{document}
\preprint {WIS-94/43/Oct-PH}
\draft
\date{\today}
\title{Nuclear transparencies for nucleons, knocked-out
under various  semi-inclusive conditions}
\author{A. S. Rinat and M.F. Taragin}
\address{Department of Particle Physics, Weizmann Institute of
         Science, Rehovot 76100, Israel}
\maketitle
\begin{abstract}
Using hadron  dynamics we  calculate nuclear transparencies  for protons,
knocked-out   in   high-$Q^2$,   semi-inclusive   reactions.    Predicted
transparencies are,  roughly half  a standard  deviation above  the NE18
data.  The  latter contain the  effects of binned proton  missing momenta
and  mass, and  of finite  detector acceptances.  In order to test
sensitivity we  compare
computed transparencies  without restrictions  and the same  with maximal
cuts for  missing momenta and the  electron energy loss.  We  find hardly
any variation, enabling a meaningful comparison with data and predictions
based   on   hadron   dynamics.     Should   discrepancies   persist   in
high-statistics data, the above may with greater confidence be attributed
to exotic components in the description of the outgoing proton.

\end{abstract}
\vspace{1cm}
\par
pacs{: 25.30 Fj, 24.10Eq}
\newpage

The following  note concerns the transparency  of a nuclear medium  for a
proton,  knocked-out  in  a  high-$Q^2$,  semi-inclusive  (SI)  reaction.
Simplest  is  the  electron-induced  SI reaction  $A(e,e'p)X$,  where  in
addition to the scattered electron, one also measures the kinematics of a
knocked-out proton.  For it, the  transparency
${\cal T}\equiv {\cal T}^{SI}$ is  the ratio  of
the  experimental  yield  and some  reference  cross  section.
Desiring it to be
a measure for  the interaction  of the exiting nucleon  with the
remaining  nuclear core, the  natural choice for  reference
cross  section  is  the  $A(e,e'p)X$ yield  where  that  interaction
is absent, i.e. the  (non-measurable) cross  section in  the Plane  Wave
Impulse Approximation  (PWIA) \cite{gar}.  Also  in use is $Z$  times the
$ep$ Mott cross section, but in that case the resulting ratio
does not tend to 1  in the above limit.

To the extent that the knocked-out particle is a hadron, conventional
nuclear dynamics
should  suffice   for  a  description   of  ${\cal T}$. However,  for
sufficiently high energies the description of the knocked-out particle is
predicted to  require sub-nucleonic
degrees of freedom, which interact  only weakly
with the nuclear  medium \cite{far}.  A feducial detection  of
the resulting color
transparency  (CT)  from  high-$Q^2$  knock-out  reactions  then  clearly
requires

i) A measurement of ${\cal T}$ under well-controlled conditions.

ii)  An  accurate   calculation  of  the  non-exotic   component  of  the
transparency.

Both requirements are hard to meet  for SI $(e,e'p)$ processes.  Data are
in general not genuinely semi-inclusive, but contain contaminations, e.g.
cuts in observables and detector acceptances \cite{gar,n18,n181}.  On the
theoretical side  the complexity of  the many-body problem stands  in the
way  of  an accurate  calculation  of  ${\cal  T}$.   In the  absence  of
deconvoluted  data,   calculations  have  sometimes  modeled   the  above
problematic   features,  relaxing   to   some  extent   the  measure   of
semi-inclusivity.  This has  led to confusion and it is  not always clear
what  is  meant  by  $'$a   standard  Glauber  calculation$'$  for  a  SI
transparency.   For clarification  we  shall use  the  approach of  Refs.
\cite{ko1,ko2} with modifications expounded in \cite{rj}.

Having defined  the transparency ${\cal T}$ and the framework for
computation, one may use the same to calculate related quantities, which
experimental conditions may force
one to study. Those become some measure for ${\cal T}$, but are not
identical to it. The spread in
outcome of  the analysis then  establishes  bounds  for the
sensitivity  of ${\cal T}$ for  various, controlled deviations from
semi-inclusivity.

Consider the  SI yield  when the electron  transfers momentum  and energy
$(q,\omega)$  to  the   target and the
knocked-out nucleon  has momenta ${\mbox{\boldmath  $p$}}$.
Having factored out the (off-shell) Mott cross section from the yield,
the  focus is  on  the remaining SI response  per nucleon \cite{moug}
\begin{eqnarray}
S^{SI}(q,\omega, {\mbox{\boldmath $ p$}})
&=&1/A\sum_n
|{\cal F}_{0n}(q)|^2\delta(\omega+\Delta_n-e_{\bmp})
\nonumber\\
{\cal F}_{0n}(q)&=&\langle\Phi^0_A|\rho_q^{\dagger}|
\Psi^{+}_{n,{\mbox{\boldmath $ p$}}}\rangle
\label{a1}
\end{eqnarray}
Here ${\cal F}_{0n}$ is the inelastic charge form factor for a transition
between the ground and an excited scattering state.  The latter describes
asymptotically a proton with mass $M$ and
total on-shell energy $e(\bmp)= (\bmp^2+M^2)^{1/2}$
and a core of $A-1$ nucleons in an excited state $n$,
separated from the ground state  by an energy $\Delta_n$.  Exploiting the
high momentum of  the exiting proton one may approximately factorize
those into states for the core and the knocked-out proton
scattered  in the field  of fixed scatterers
\begin{eqnarray}
\Psi^{+}_{n,\bmp}\approx \Phi^{A-1}_n\psi^+_{\bmp}
\label{a2}
\end{eqnarray}
Substitution of (2) into (1) and subsequent application of closure leads
to \cite{rj,gr2} ($\bms=\bmz-\bmzp$)
\begin{mathletters}
\label{a3}
\begin{eqnarray}
S^{SI}(q,y_0,\bmp)&=&\frac{m}{q}\delta(y_0-p_z) \int d{\bmr}_1\int d\bms
e^{ i{\mbox{\boldmath $ ps$}} }
\rho_1(\bmr_1,\bmr_1') \tilde R(q,\bmr_1',s_z)
\label{a3a}\\
\tilde R(q,\bmr_1',s_z)&=& \bigg ( \prod_j \int d{\bmr}_j\bigg )
\frac{\rho_A({\bmr}_1,{\bmr}_1';{\bmr}_j)}{\rho_1({\bmr}_1,{\bmr}_1')}
\prod_{l\ge 2}
\bigg (1+\gamma(q,{\bmr}_1-{\bmr}_l;{\bms})\bigg )
\label{a3b}
\end{eqnarray}
\end{mathletters}
The energy loss $\omega$ in (\ref{a1})
has in (\ref{a3a}) been replaced by a scaling variable
\begin{eqnarray}
y_0&\approx&\bar{y}_0\bigg[1-\frac{1}{2A}\bigg(1+\frac{q}{\bar{y}_0}
\bigg)^{-1}\bigg (1+\frac{\omega-\langle\Delta\rangle}{M}\bigg)\bigg]
\nonumber\\
\bar{y}_0&=&-q+\sqrt{2M(\omega-\langle \Delta\rangle)+(\omega-
\langle \Delta\rangle)^2},
\label {a4}
\end{eqnarray}
with  $\langle\Delta\rangle$, an average nucleon
separation  energy  for the  nuclear  species  considered.

For any  inclusive  process  the nuclear  input required  in
(\ref{a3b})  is the  nuclear  density matrix  $\rho_A$,  diagonal in  all
coordinates except in the one  of the struck particle ($'$1$'$). In
contradistinction,
the expression for $\gamma$ in  (\ref{a3b}) which describes  Final State
Interactions (FSI) depends  on the type of  the SI process as  well as on
the approximation used in the evaluation. The
induced high-energy rescattering of the knocked-out nucleon from the
core is for all processes  routinely described by Glauber theory
\cite{gl}.

One thus finds from (\ref{a1}), (\ref{a2}) for the $'$unrestricted$'$ SI
process (i.e.  without binning of data) \cite{ko1}
\begin{eqnarray}
\gamma(q,\bmr,s)=\bigg (1+\Gamma_q^{off}(b,z)\bigg )\bigg(1+
\lbrack\Gamma_q^{off}(b,z+s)\rbrack^*\bigg ),
\label{a5}
\end{eqnarray}
where for short-range $NN$ interactions $V$ one may approximately relate
the elastic off-shell profile function $\Gamma^{off}$ in (\ref{a5}) to
its corresponding on-shell analog $\Gamma$. With $v_q=e(q)/q$
\begin{eqnarray}
\Gamma_q^{off}(b,z)&\equiv&
{\rm exp}[-(i/v_q)\int_z^{\infty} d\zeta V(b,\zeta)]-1
\approx{\rm exp}[-(i/v_q)\theta(-z)\int_{-\infty}^{\infty} d\zeta
V(b,\zeta)]-1
\nonumber\\
&=&\theta(-z)\Gamma_q(b)=\theta(-z) \bigg (
{\rm exp}[-(i/v_q)\int_{-\infty}^{\infty} d\zeta
V(b,\zeta)]-1\bigg )
\label{a6}
\end{eqnarray}
For the on-shell profile
$\Gamma$ we use the standard  parametrization
\begin{eqnarray}
\Gamma_q(b)=e^{i\chi_q(b)}-1\approx
-(\sigma_q^{tot}/2)(1-i\tau_q)A(q,\bmb),
\label{a7}
\end{eqnarray}
with $\tau$, the ratio of real and imaginary parts of the forward
$NN$ elastic scattering amplitude, and  $A(q,\bmb)$
accounting for the  range  of  $\Gamma_q(b)$;  zero-range
corresponds to $A(q,\bmb)\to\delta^2(\bmb)$.
Inelastic contributions to elastic $pp$  and $pn$ scattering are
accounted for by the imaginary part of the eikonal
phase $\chi$ in (\ref{a7}) and are
in the impact  parameter representation
related  to the  $'$partial$'$ inelastic cross section
$\sigma^{p,inel}(b)$
\begin{eqnarray}
\sigma_q^{p,inel}(b)
&\equiv& 1-e^{-2{\rm Im}\chi_q(b)} \approx A(q,\bmb)\sigma_q^{inel}
\nonumber\\
A(q,\bmb)&\approx& \frac {(Q_q^0)^2}{4\pi}{\rm exp}\lbrack-(bQ_q^0/2)^2
\rbrack
\nonumber\\
\sigma_q^{inel}&=&\sigma_q^{tot}-\sigma_q^{el}
\label{a8}
\end{eqnarray}
Using (\ref{a7}), (\ref{a8}) one shows that $\gamma$ in Eq. (\ref{a5})
becomes
\begin{eqnarray}
\gamma(q,\bmr,s)&=& -A(b)\bigg\lbrack \theta(-z)\langle \sigma^{inel}
\rangle+\theta(z)\theta(s-z)\bigg\langle \frac
{1+i\tau}{2}\sigma^{tot}\bigg \rangle \bigg\rbrack
\nonumber\\
\langle \beta\sigma\rangle&=&(A-1)^{-1}\bigg \lbrack(Z-1)\beta^{pp}
\sigma^{pp}+ N\beta^{pn}\sigma^{pn}\bigg \rbrack,
\label{a9}
\end{eqnarray}
where we defined averaged (weighted) $pN$ cross sections.

Eqs. (\ref{a3}) describe in principle
FSI to all orders in $\gamma$, but for large momentum
transfers it suffices to retain terms up to second order in $\gamma$.
Those involve $\rho_n, n\le 3$  \cite{rt,grs}
\begin{eqnarray}
\rho_n(\bmr_1,\bmr_1';{\bmr}_j)
&=& \rho_1({\mbox{\boldmath $ r_1,r'_1$}})\bigg \lbrack
\prod_{j\ge 2}^{n-1} \rho({\bmr}_j) \bigg \rbrack
\zeta_n({\mbox{\boldmath $ r$}}_1,
{\mbox{\boldmath $ r$}}_j;{\mbox{\boldmath $ s$}})
\nonumber\\
\zeta_2({\mbox{\boldmath $ r$}}_1,
{\mbox{\boldmath $ r$}}_2;{\mbox{\boldmath $ s$}})
&\approx& \sqrt {g(\bmr_1-\bmr_2) g(\bmr_1-\bmr_2+s_z\hat{\bmq})}
\nonumber\\
\zeta_n(\bmr_1,..,\bmr_n,\bms)&\approx&\prod_{j\le k}
\zeta_2(\bmr_j,\bmr_k,s_z)
\label{a10}
\end{eqnarray}
In all $\rho_n$ we factored out the non-diagonal
single-particle  density
$\rho_1(\bmz,\bmzp)$ which may approximately be parametrized as
\cite{vaut}
(${\mbox{\boldmath $ S$}}=(\bmz +\bmzp)/2;
\bms=\bmz-\bmzp$)
\begin{eqnarray}
\rho_1(\bmz,\bmzp)&\approx&\rho(\mbox{\boldmath $ S$)}\int d\mbox
{\boldmath $ S$}'\rho_1(\bms,\mbox{\boldmath $ S$}')\approx
\rho(\mbox{\boldmath $ r$}_1)
\Sigma(\bms)
\nonumber\\
\Sigma(\bms)&=&\int \frac {d^3\bmp}{(2\pi)^3} n(\bmp)e^{-i\bmp\bms}
\label{a11}
\end{eqnarray}
and where $\rho(\bmr)=\rho_1(\bmr,\bmr)$ is the single
nucleon density. Next
one encounters in the expression for $\rho_2$ in (\ref{a10})
a non-diagonal pair-distribution function $\zeta_2$, approximated
there by means of the diagonal pair-distribution function $g$
\cite{grs}. The non-diagonal $n$-body distribution  function $\zeta_n$
in (\ref{a10}) is, in the  independent pair approximation
expressed as a  product of pair analogs $\zeta_2$.
Using (\ref{a10}) and  (\ref{a11}), Eqs. (\ref{a3}) become
\begin{mathletters}
\label{a12}
\begin{eqnarray}
S^{SI}(q,y_0,\bmp)&=&\frac{m}{q}\delta(y_0-p_z) \int d\bms
e^{ i{\mbox{\boldmath $ ps$}} }\Sigma(|\bms|)
\int d{\mbox{\boldmath $r_1$}}\rho(\bmr_1)
\tilde R(q,\mbox{\boldmath $r_1$},s_z)
\nonumber\\
&=&\frac{m}{q}\delta(y_0-p_z) \int d\bms
e^{i{\mbox{\boldmath $ ps$}}}\Sigma(|\bms|) {\tilde G}(q,s_z)
\label{a12a}\\
{\tilde G}(q,s_z)&=&\int d{\mbox{\boldmath $r_1$}}\rho(\bmr_1)
\tilde R(q,\mbox{\boldmath $r_1$},s_z)
\label{a12b}\\
S^{SI,PWIA}(q,y_0,\bmp)&=&\frac{m}{q}\delta(y_0-p_z) \int d\bms
e^{ i{\mbox{\boldmath $ ps$}} }\Sigma(|\bms|)
\int d\bmr_1\rho(\bmr_1)= (m/q)\delta(y_0-p_z) n(\bmp)
\label{a12c}
\end{eqnarray}
\end{mathletters}
Treating the retained FSI terms in the first cumulant approximation,
one finds for $A\gg 1$
\begin{mathletters}
\label{a13}
\begin{eqnarray}
\tilde R&=&e^{\Omega}=e^{\Omega_2+\Omega_3+...}
\label{a13a}\\
\Omega_2(q,{\bmr}_1;s)&=&(A-1)\int d{\bmr}_2
 \rho(r_2)\gamma(q,{\bmr}_1-{\bmr}_2,s)
 \zeta_2({\bmr}_1- {\bmr}_2,s)
\label{a13b}\\
\Omega_3(q,{\bmr}_1;s)&=&\frac{A-1}{2}
\int \int d{\bmr}_2 d{\bmr}_3
 \rho(r_2)\rho(r_3)\gamma(q,{\bmr}_1-{\bmr}_2,s)
 \gamma(q,{\bmr}_1-{\bmr}_3,s)\zeta_3({\bmr}_1,\bmr_2,\bmr_3,s)
\label{a13c}
\end{eqnarray}
\end{mathletters}
We shall later on return to the ternary collision terms above and
discuss first the binary collision part in (\ref{a13}) \cite{foot1}
\begin{mathletters}
\label{a14}
\begin{eqnarray}
\tilde R_2&=&\tilde R_2^{tot}\tilde R_2^{inel}
\nonumber\\
\tilde R_2^{tot}(q;\bmr_1,s_z)
&\approx&\hbox{\rm exp}\bigg \lbrack
-(A-1)\bigg \langle \frac {1+i\tau_q}{2} \sigma^{tot}_q\bigg \rangle
\nonumber\\
&&\int d^2\bmb A(b) \int_0^{s_z}dz
\rho(\bmb_1-\bmb,z_1-z)\zeta_2(\bmb,z,s_z)\bigg]
\label{a14a}\\
\tilde R_2^{inel}(q;{\mbox{\boldmath $ r$}}_1,s_z)
&\approx&\hbox{\rm exp}\bigg \lbrack
-(A-1)
\bigg\langle\sigma^{inel}_q\bigg\rangle \int d^2\bmb A(b)
\nonumber\\
&&\bigg( \int_{-\infty}^0dz \rho(\bmb_1-\bmb,z_1-z)\zeta_2(\bmb,z,s_z)
\bigg \rbrack
\label{a14b}
\end{eqnarray}
\end{mathletters}

We note  that the  results Eq.  (\ref{a12})-(\ref{a14}) contain
measurable  quantities  without reference  to an  underlying potential
method.   It seems then  reasonable to dissociate  the outcome from that
model and to postulate its validity also in the high-$Q^2$ regime.

We start with Eqs. (\ref{a12a}), (\ref{a12c}) for
the $'$unrestricted$'$ nuclear transparency, without any binning or
correction for detector acceptance
\begin{mathletters}
\label{a15}
\begin{eqnarray}
{\cal T}(q,\bmp)
&\equiv & \frac{S^{SI}(q,y_0,\bmp)}
{S^{SI,PWIA}(q,y_0,\bmp)}=\frac{ \int d\bms
e^{i{\mbox{\boldmath $ ps$}}}\Sigma(|\bms|){\tilde G}(q,s_z)}
{ \int d\bms e^{i{\mbox{\boldmath $ ps$}}}\Sigma(|\bms|)}
\label{a15a}\\
&=& [n(p)]^{-1}\int d\bms
e^{i{\mbox{\boldmath $ ps$}}}\Sigma(|\bms|){\tilde G}(q,s_z)
={\lbrack n(p)\rbrack}^{-1}\int \frac {dp_z'}{2\pi}
n({\mbox{\boldmath $p_{\perp}$}},p_z-p_z') G(q;p_z')
\label{a15b}
\end{eqnarray}
\end{mathletters}
${\cal T}$ is from (1) expected to be a function of $q$, the energy loss
$\omega$ or  $y_0$, Eq. (\ref{a4}), and the momentum of
the knocked-out proton momentum \bmp. The fact that (\ref{a15}) does not
appear to depend  on $y_0$ is  an artifact  of closure. Without its
application one is lead to  (\ref{a15b}) with non-canceling
single-nucleon spectral functions instead of momentum distributions.

Next  we  consider  conditions,   intermediate  between  semi  and  total
inclusive  scattering and  consider first  cuts on  the missing  momentum
$\bmp_m=\bmp-\bmq$.   For fixed  $\bmq$ those  are  the same  as cuts  on
$\bmp$.  We  thus integrate $both$ SI yields or responses  (\ref{a3}) in
the ratio (\ref{a15a})  over the momentum $\bmp$ of  the outgoing proton,
with the result
\begin{mathletters}
\label{a16}
\begin{eqnarray}
{\cal T}^{P}(q,y_0)
&\equiv&  \frac {\int d\bmp S^{SI}(q,y_0,\bmp)}
{\int d\bmp S^{SI,PWIA}(q,y_0,\bmp)}
=\frac {\int ds_z
e^{iy_0s_z }\Sigma(s_z)\tilde G(q,s_z)}
{ \int d{s_z} e^{iy_0s_z }\Sigma(s_z)}
\label{a16a}\\
&=& \int \frac {dp'}{(2\pi)} F_0(y_0-p')G(q,p'),
\label{a16b}
\end{eqnarray}
\end{mathletters}
and where we used the asymptotic limit of the NR total inclusive response
\cite{grs}
\begin{eqnarray}
F_0(y_0)=\lim_{q\to\infty}\frac {m}{q} S^{TI}(q,y_0)=
\int \frac{d\bmP}{(2\pi)^3}n(\bmP)\delta (y_0-P_z)
\label{a17}
\end{eqnarray}
After the remark following Eq.  (\ref{a15}) one observes the reappearance
of  the energy  loss  through  $y_0$  in  ${\cal T}^P$,  Eq. (\ref{a16}).
Notice that the response $S^{SI,PWIA}$ in the ratio in (\ref{a16a}), when
integrated  over the  the proton  momentum and  energy loss  is 1,  which
implies $Z\sigma^{Mott}$ for the  corresponding  reference
cross section.  This does
not contradict  the original choice of  PWIA yields as reference  for the
definition of  ${\cal T}$: it emerges if in (\ref{a15a}) one consistently
applies the same cuts in  actual and reference yields.

Finally we consider the transparency  when the cross sections in  the
ratio (\ref{a16}) are in addition
integrated over the energy loss of the electron. From
Eqs. (\ref{a3}) and (\ref{a8}))  one readily finds \cite{foot2}
\begin{mathletters}
\label{a18}
\begin{eqnarray}
{\cal T}^{PE}(q) &\approx& \tilde{G}(q,0)
= \int d{\bmr}_1\rho({\bmr}_1) \tilde{R}(q,{\bmr}_1,0)
\label{a18a}\\
&=&\int d{\bmr}_1 \rho({\bmr}_1)
\hbox{\rm exp}\lbrack -(A-1)\langle \sigma^{inel}_q\rangle
\int d^2\bmb A(\bmb)\int^{\infty}_0 dz \rho(\bmb_1-\bmb,z_1-z)
g(b,z)\rbrack
\label{a18b}\\
&\approx&
\lbrack {(A-1) \langle g\rangle \langle\sigma_q^{inel}}\rangle
\rbrack ^{-1}
\int d{\mbox{\boldmath $ b$}}\bigg \lbrack
1-\hbox{\rm exp}\bigg (-(A-1) \langle g\rangle
\langle\sigma^{inel}_q\rangle \lbrace t_2(b)+t_3(b)+..\rbrace
\bigg ) \bigg \rbrack
\label{a18c}
\end{eqnarray}
\end{mathletters}
Eq. (\ref{a18c}) holds only in the  0-range interaction limit and
$t_2(b)=\int^{\infty}_{-\infty}\rho(\bmb,z')$  is the standard
thickness  function resulting from binary FSI. $t_3(b)$ there
is due to ternary contributions, to be discussed shortly.

Although the  actual derivation above leads to
$\sigma^{inel}$,  one occasionally  meets
expressions for ${\cal T}$ with $\sigma^{inel}\to
\sigma^{tot}$ \cite{pp} where in particular the binary part
\begin{mathletters}
\label{a19}
\begin{eqnarray}
{\cal T}_2^{PP}(q) &\to&
\int d{\mbox{\boldmath $ r$}}_1 \rho({\mbox{\boldmath $ r$}}_1)
\hbox{\rm exp}\lbrack (-(A-1)\langle\sigma^{tot}_q\rangle
\int^{\infty}_0 dz \rho(b_1,z_1-z) g(0,z)\rbrack
\label{a19a}\\
&\approx&
\lbrack {(A-1) \langle g\rangle \langle\sigma_q^{tot}}\rangle\rbrack^{-1}
\int d^2\bmb \bigg \lbrack
1-\hbox{\rm exp}\bigg (-(A-1)\langle g\rangle \langle\sigma^{tot} \rangle
t_2(b) \bigg ) \bigg \rbrack,
\label{a19b}
\end{eqnarray}
\end{mathletters}
is reminiscent of the transparency for an  elastically scattered
proton. The above has for instance been used by Benhar et al, who
extended a previous description  of the totally inclusive $(e,e')$
process \cite{om} and  reached  ${\cal  T}^{PP}$, Eq. (\ref{a19a})
\cite{om1}.  The  same holds  for Frenkel,
Frati and Walet \cite{ffw}  and  the criticism voiced in \cite{ko1,ko2}
and in particular in \cite{rj} pertains to it as well.
Finally we mention work by the Rome-Perugia group, who use a
spectral function beyond the mean-field approximation, but do not seem to
go beyond the PWIA \cite{cl1}.

We now discuss Eq. (\ref{a18c}) in some detail.
Its binary collision version (with
ternary and higher order corrections $t_3+...\to 0$) resulted when
Kohama $et\,al$  \cite{ko1,ko2} applied
unnecessarily strong approximations in an evaluation of the
$'$unrestricted$'$  transparency ${\cal T}$, Eq. (\ref{a15}).

Alternatively, desiring  to simplify  matters, Nikolaev $et\,al$
just replaced  ${\cal T}\to {\cal T}^{PE}$, obtained from cross sections
integrated over missing  momentum and energy-loss \cite{nnn}. Those
authors considered binary ($'$hole$'$) as well as ternary
($'$spectator$'$) contributions  with the scattering operator
(\ref{a18a}), appropriate to ${\cal T}^{PE}$, i.e.
(\ref{a5}) with $s\to  0$: The
integration over the energy loss eliminates the retardation in the
propagation of the density disturbance. As a consequence static
(diagonal) quantities  replace everywhere
non-diagonal ones, e.g. $\zeta_2\to g$, etc.

Instead of $g$ Nikolaev $et\,al$ prefer in (\ref{a10})
the use  of the pair-correlation function $C=1-g$, thus
\begin{mathletters}
\label{a20}
\begin{eqnarray}
g_3(1,2,3)&\equiv&\zeta_3(1,2,3,0)\approx g(1,2)g(1,3)g(2,3)
\label{a20a}\\
&\approx& C(1,2)C(1,3)-C(2,3)
\label{a20b}
\end{eqnarray}
\end{mathletters}
where for brevity we wrote $g(12)=g(\bmr_1,\bmr_2)$, etc.
Making the additional approximations
$g=-C$ and the second part of (\ref{a20b}) in Eqs. (\ref{a13b}),
respectively (\ref{a13c}) for $s=0$, those become
\begin{mathletters}
\label{a21}
\begin{eqnarray}
\Omega_2^{PE,N}(q,{\bmr}_1)&\approx&(A-1)\int d{\bmr}_2
\rho(r_2)\gamma^{PE}(q,{\bmr}_1-{\bmr}_2)
C({\bmr}_1,{\bmr}_2)
\label{a21a}\\
\Omega_3^{PE,N}(q,{\bmr}_1)&\approx&\frac{A-1}{2}
\int \int d{\bmr}_2 d{\bmr}_3
 \rho(r_2)\rho(r_3)\gamma^{PE}(q,{\bmr}_1-{\bmr}_2)
 \gamma^{PE}(q,{\bmr}_1-{\bmr}_3)C(\bmr_2,\bmr_3)
\label{a21b}
\end{eqnarray}
\end{mathletters}
Next one shows that
the product $CC$ in (\ref{a20b}) equals $(1/2)\bigg
(\Omega^{PE,N}\bigg )^2$ and similar approximations to higher order
densities produce terms of the same order
in the expansion of the first cumulant
of (\ref{a21a}). Nikolaev $et\,al$ thus reach
\begin{eqnarray}
R^N&=&{\rm exp} \bigg (\Omega_2^N(\lbrack C \rbrack)+\Omega_3^N+..\bigg )
\nonumber\\
\Omega_2^N\lbrack C \rbrack)&=&\Omega_2\lbrack g \rbrack)
\label{a22}
\end{eqnarray}
In view of the replacement $g\to -C$ in the actual expression
Eq. (\ref{a13b}) for $\Omega_2$, that approximation misses contributions
to $\Omega^2$ linear in  $\gamma$. Likewise, there are
ternary FSI contributions quadratic in $\gamma$ which are not contained
in either $1/2(\Omega^N)^2$ or in (\ref{a21b}).

Comparison of numerical results reflects the above shortcomings.
Nikolaev $et\,al$ report that the first cumulant of binary as well as
ternary FSI contributions  Eqs. (\ref{a21})
are $+$ (5--6)$\%$, respectively -(2--3)$\%$ of the PWIA result. However,
the former fraction, computed from
(\ref{a13b}) with $C\to -g$  is much larger,
namely 15--20 $\%$ \cite{pp,rj}. Even assuming (\ref{a22b}) to be a
measure for ternary FSI, those cut the binary part by
only $\approx$ 15$\%$  and not by $\approx$50$\%$.
Clearly, in view of the approximations involved, it seems
safe to disregard ternary FSI altogether.

We now present some numerical results  for the transparencies of a number
of  nuclear  species  under  NE18 kinematics  \cite{n18,n181},  i.e.  for
$Q^2$=1.0, 3.0,  5.0 and 6.7 GeV$^2$,  ${\bmp}\approx |\bmp|\hat q$.
For  the evaluation  of the  predictions Eqs.
(\ref{a12})-(\ref{a18}) one needs nucleon densities $\rho(r)$
and pair distribution functions $g$ (both taken to be the
same for protons and neutrons)  and  dynamical input.
Measured or interpolated  values for representative $NN$
scattering parameters are given in  Table I.
A glance at the entries suffices for one prediction:
In the region of the NE18 experiment transparencies  should,
for any given nuclear species decrease as function of $Q^2$ and reach a
plateau beyond $Q^2\approx $ 1.5 GeV$^2$. Any significant deviation
from that prediction  cannot be accounted for by standard hadron physics.

In Figs.  1a,b,c we entered  for C,  Fe and Au  as function of  $Q^2$ the
undeconvoluted NE18 data \cite{n181} and those are compared with the
following predictions (none contain corrections for finite solid-angle
detector acceptance).

i)  Unrestricted transparencies  ${\cal T}$,  Eq. (\ref{a15}),  caused by
binary collisions, which  in turn are generated by finite,  as well as by
zero-range interactions.

ii) Results for ${\cal T}^P$  (\ref{a16}), when cross sections have been
integrated over  missing momentum.  Compared  to ${\cal T}$,  changes are
$\leq  2$ $\%$,  independent of  the nuclear  species or  $Q^2$, and  are
insensitive to any reasonable choice for the average separation energy.

iii) ${\cal T}^{PE}$, Eq. (\ref{a18a}),  obtained from cross sections, in
addition integrated over the electron  energy loss.  Relative changes now
amount to  (4-6)$\%$.  We  checked that  in all  cases only  tiny changes
occur if one discards the non-negligible $\tau$ (Table I).
Out of all parameters studied, only the  $NN$  interaction
range affects predictions.

iv) Sizably smaller transparencies ${\cal T}^{PP}$,
Eq. (\ref{a19b}), result from  Eq.  (\ref{a19a}),  where total  cross
sections  replace the  smaller  inelastic ones.

We conclude:

1) For all targets and all $Q^2$, the most complete  calculations using
binary FSI produce transparencies, roughly  half a standard deviation
higher than the data. Lacking data with better statistics, it seems
premature to ascribe the current discrepancies to color transparency,
whose onset is anyhow  not expected to occur for $Q^2$ as low as
$\approx$ 1-1.5 GeV$^2$. Ternary FSI contributions decrease the above
results by only a few  percent.

2) There are only modest changes in ${\cal}T$ due to acceptable ranges
in input parameters. Likewise ${\cal}T$ is  only marginally  sensitive to
$'$maximal$'$ cuts and it seems reasonable to assume the
same  for actual, more restricted cuts.

3) Having demonstrated the stability of
calculations based on conventional hadron dynamics, firm
conclusions can be drawn regarding the existence of exotic
parts in ${\cal T}$, should discrepancies with high-statisitic data
persist. Presently one can only say that no conventional
hadron theory  can  predict  the  rise  in  the
undeconvoluted NE18 data towards the largest measured $Q^2$ points for
all targets, or the relatively flat carbon data for the lowest $Q^2$.

\newpage

{\bf  Figure Captions.}

Fig. 1a.  Transparencies of  C the  passage of  a proton,
knocked out in a semi-inclusive  $A(e,e'p)X$ reaction for NE18 kinematics
as function of $Q^2$.  Data are from O'Neill $et\,al$ \cite{n181} and
differ slightly from previously published  data by
Makins $et\,al$ \cite{n18}.
The drawn, dotted and long dashed lines correspond
to  respectively  ${\cal  T}$,  ${\cal  T}^P$,  ${\cal  T}^{PE}$  ,  Eqs.
(\ref{a15}), (\ref{a16}) and (\ref{a18c}).  The dot-dashed line gives the
0-range interaction result  for ${\cal T}$.  Short dashes  are for
${\cal T}^{PP}$, Eq. (\ref{a19a}).

Fig. 1b. Same as Fig. 1a for Fe. Data are from O'Neill $et\,al$
\cite{n181}.

Fig. 1c. Same as Fig. 1b for Au.

\vspace{2cm}
\par

Table I.  Partly interpolated $pp$ and $pn$ elastic scattering parameters
\cite{dat}.
\vspace{1cm}
\setlength{\textwidth}{6.3in}
\begin{tabular}{|rr|rr|rr|rr|rr|} \hline

$q$ & $Q^2$ & $\sigma^{tot}_{pp}$ & $\sigma^{inel}_{pp}$&
$\sigma^{tot}_{pn}$ & $\sigma^{inel}_{pn}$ & $\tau_{pp}$ & $\tau_{pn}$&
$(Q_0)_{pp}$ & $(Q_0)_{pn}$ \\
(GeV) & (GeV$^2$)&(mb) &(mb) & (mb) & (mb)&     &      &
 (GeV) & (GeV)\\ \hline\hline
 1.20 & 1.04 & 36.6 & 12.6 & 36 & 8 & 0.30 & -0.25 & 0.84 & 0.84 \\
\hline
 2.40 & 3.0 & 46 & 26 & 42 & 28 & -0.35 & -0.40 & 0.59 & 0.59 \\
\hline
 3.54 & 5.0 & 42.4 & 27.1 & 44.0 & 27.1& -0.37 & -0.45 & 0.56 &
0.56 \\ \hline
 4.48 & 6.7 & 42.4 & 27.5 & 44.0 & 27.5& -0.37 & -0.45 & 0.56 &
0.56 \\ \hline
\end{tabular}

\begin{references}
\bibitem{foot2} The proton momentum, energy-loss integrated cross section
is related, but not identical, to the total inelastic cross section
\cite{mg}. The latter includes all $A$-particle final states,
disregarding production processes, while the former builds those
from  $A$-1 particle  and single proton scattering  states (see also
\cite{mm} for a semi-empirical derivation of (\ref{a18b})).
\end{references}
\end{document}